# Interferometric detection of mode splitting for whispering gallery mode biosensors

Joachim Knittel,[*] Terry G. McRae, Kwan H. Lee, and Warwick P. Bowen

*School of Mathematics and Physics, University of Queensland, Brisbane, QLD 4072, Australia*

[*]*Corresponding author: j.knittel@uq.edu.au*

Sensors based on whispering gallery mode resonators can detect single nanoparticles and even single molecules. Particles attaching to the resonator induce a doublet in the transmission spectrum which provides a self-referenced detection signal. However, in practice this spectral feature is often obscured by the width of the resonance line which hides the doublet structure. This happens particularly in liquid environments that reduce the effective Q factor of the resonator. In this paper we demonstrate an interferometric set-up that allows the direct detection of the hidden doublet and thus provides a pathway for developing practical sensor applications.

Whispering-gallery mode (WGM) based sensors are able to detect single molecules and nanoparticles in liquid and gaseous environments and constitute an ultra-sensitive, on-chip platform for label-free biological, chemical and medical sensing applications [1]. Specific detection can be achieved by functionalizing the sensor surface, i.e. with anti-bodies, that exclusively allow binding of distinctive target molecules [2]. The presence of a particle is detected via a resonance frequency shift that is proportional to the polarizability of the detected particle. The polarizability can then be used to estimate the size of the particle provided that its refractive index is known [1]. Generally the frequency shift is deduced from the transmission spectrum that is obtained with a laser whose frequency is periodically scanned over the resonance.

To reduce measurement noise introduced by sources, such as laser wavelength, detector noise, and temperature drifts of the resonator and the surrounding medium, a self-referencing detection method based on mode splitting was recently demonstrated [3]. This method relies on the scattering induced mode-splitting of the originally degenerate clockwise (CW) and counter-clockwise (CCW) propagating modes of the WGM resonator. The splitting, i.e. the frequency separation of the doublet lines, depends on the coupling rate $g$ of the CW and CCW propagating modes and has the value $2g$. From the splitting and the two linewidths of the doublet the size of the particle can be estimated with high fidelity.

Generally, the splitting is interfered from the spectrum around the resonance lines, by fitting a double Lorentzian function to it [3]. To achieve an accurate fit the splitting should be larger than the linewidth of the resonance. In other words, the sum of all losses should not reduce the quality factor $Q$ of the resonator below $\omega_c/(2g)$, where $\omega_c$ is the resonance frequency of the cavity. Otherwise, the mode splitting feature is hidden behind the spectrally broadened resonance and prevents the accurate measurement of the splitting from the spectra. For example the splitting caused by the relatively large single virus (InfA A/PR/8/34) is on the order of 30MHz [4]. To resolve the doublet directly a $Q$ factor of at least $10^7$ is required. This can be routinely achieved in atmosphere under a laboratory environment. However, practical applications require the sensor to be operated within a liquid medium, where these ultra-high $Q$ factors are much harder to achieve, presenting a severe limit on the number of possible applications. For example in water the $Q$ factor of a WGM sensor operating at wavelength of 670nm is typically below $10^6$ [5]. A further constraint of current systems is the requirement of scanning the laser around the resonance line and thus the bandwidth is typically limited to about 1 kHz by the piezoelectric element that scans the frequency of the laser [2] and the data processing speed [3]. However, a high bandwidth is essential to observe fast molecular events such as diffusion processes.

In this letter we resolve these problems implementing an interferometric set-up that allows the splitting to be quantified in real time with a bandwidth beyond 10 MHz even when the doublet is obscured. This provides the capacity to sense small, single unlabeled molecules and may be an enabling technology to investigate single-molecular processes, protein folding [6] and to observe motor proteins at work [7].

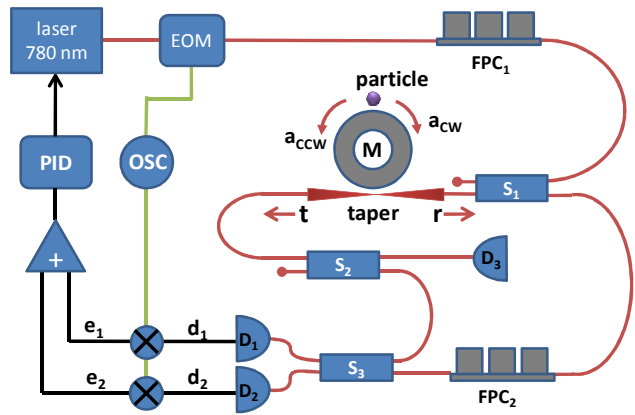

Fig. 1: Schematic of the fiber interferometer with 50/50 beamsplitter S, detector D, Oscillator OSC, fiber polarization controllers FPC and phase modulator EOM. The microtorroid M is mounted on a 3 axis piezoelectric translation stage. The splitting generated by a particle is detected via the interference signals $d_1$ and $d_2$.

A schematic of our interferometric set-up is shown in Fig. 1. The light from a tunable single-mode laser is evanescently coupled into a WGM resonator via a fiber taper after passing the 50:50 fiber beamsplitter $S_1$. The coupling strength between taper and resonator can be adjusted by changing the position of the taper with a 3-axis piezoelectric stage. In the resonator the coupled light generates a counter-clockwise (CCW) propagating mode with amplitude $a_{ccw}$. A particle that attaches to the surface of the cavity and is small compared to the wavelength will scatter light into the originally degenerate and unpopulated clockwise (CW) propagating mode with amplitude $a_{cw}$ [8].

To measure the splitting we collect the fields leaving the cavity via the fiber taper in forwards and backwards propagation direction, corresponding to the CW and CCW cavity modes, respectively. After passing through the beamsplitters $S_1$ and $S_2$, the fields are recombined with a third 50:50 beamsplitter $S_3$ to generate two interference signals $d_1$ and $d_2$ that are detected with the photo detectors $D_1$ and $D_2$. These interference signals correspond to the standing modes in the cavity, as we will explain in the following section. The additional splitter $S_2$ is introduced to ensure that the fields leaving the cavity in forward and backward direction both experience identical losses, as splitter $S_1$ is required to couple the light from the laser into the resonator.

The equations governing the field distribution in the WGM resonator have already been derived in several instances and are reproduced here for convenience [3, 8]. It is natural to transfer the CW and CCW modes into a normal mode basis, where the normal modes $a_+$ and $a_-$ correspond to two standing waves in the resonator and are given by:

$$a_+ = (a_{CW} + a_{CCW})/\sqrt{2} \quad \text{and} \quad a_- = (a_{CW} - a_{CCW})/\sqrt{2} \qquad (1)$$

The electric field amplitude of the normal modes $a_+$ and $a_-$ in the steady-state regime is given by:

$$a_+ = \frac{\sqrt{2\kappa_1}\, a_{in}}{\kappa_0 + \kappa_1 + 2\Gamma_R - 2i(\Delta - 2g)} \quad \text{and} \quad a_- = \frac{\sqrt{2\kappa_1}\, a_{in}}{\kappa_0 + \kappa_1 - 2i\Delta} \qquad (2)$$

where $\Delta = \omega - \omega_c$ denotes the laser frequency detuning from the unperturbed resonance frequency $\omega_c$. The damping rate $\kappa_0$ describes the intrinsic damping due to material and radiation losses, and $\kappa_1$ is the fiber taper resonator coupling rate. These two coefficients are related to the loaded Q factor of the resonator by $\kappa_0+\kappa_1= \omega_c/Q$. The term $a_{in}$ corresponds to the normalized amplitude of the input intensity. Note that the resonance frequencies, i.e. the $\omega$ corresponding to the maximum mode amplitude $a_+$ and $a_-$, differ by $2g$ and that the linewidth of the frequency shifted mode is wider due to the additional loss rate $\Gamma_R$ caused by the scattering particle. The fact the particle neither shifts nor broadens the resonance line of normal mode $a_-$ implies that the electric field of mode $a_-$ is zero at the position of the particle.

As it is impractical to measure the amplitudes of the two standing waves directly, we derived them via the $a_{cw}$ and $a_{ccw}$ amplitudes of the CW and CCW propagating modes. These modes are available through the forward transmitted fields $t$ and the backwards scattered fields $r$ at the two ends of the taper respectively (see Fig. 1), and are given by

$$t = a_{in} + \sqrt{\kappa_1} a_{ccw} \quad \text{and} \quad r = \sqrt{\kappa_1} a_{cw} \qquad (3)$$

The novelty of our set-up is that the beams $t$ and $r$ are combined and interfere at splitter $S_3$ prior to detection on detectors $D_1$ and $D_2$. The amplitudes at the two detectors are then

$$d_1 = \frac{1}{2}(a_{in} + \sqrt{\kappa_1} a_{ccw} + e^{i\phi} \sqrt{\kappa_1} a_{cw}) \quad \text{and} \quad d_2 = \frac{1}{2}(a_{in} + \sqrt{\kappa_1} a_{ccw} - e^{i\phi} \sqrt{\kappa_1} a_{cw}) \qquad (4)$$

where $\phi$ is a phase shift introduced to take account of the optical path length difference for the fields $t$ and $r$. Apart from a constant offset and a scaling factor, when a phase shift of $\phi = 1$ or $\pi$ is selected, the detector signals correspond to the amplitudes of the normal modes $a_+$ and $a_-$, as can be seen by comparison with eq. (1). A straight forward calculation using the preceding equations (1-4) yields the normalized amplitude of the light on the detectors introducing:

$$d_1 = \frac{1}{4}\left(1 - \frac{2\kappa_1}{2\Gamma_R + \kappa_0 + \kappa_1 + 2i(\Delta - 2g)}\right) \quad \text{and} \quad d_2 = \frac{1}{4}\left(1 - \frac{2\kappa_1}{\kappa_0 + \kappa_1 + 2i\Delta}\right) \qquad (5)$$

Assuming weak coupling of the taper to the cavity, i.e. $\kappa_1 < 2\Gamma_R + \kappa_0$, the spectra of both detector signals have only a single minima which correspond respectively to the resonance frequency of the normal modes $a_+$ and $a_-$. Furthermore, the width of the spectral peak in $d_1$ is affected by the scattering loss $\Gamma_R$, while that of $d_2$ is not. Thus, these two signals allow both the splitting and the scattering loss to be determined, allowing both high sensitive single particle detection and a precise measurement of the particle radius even in the regime where the cavity loss rate is much lower than the splitting.

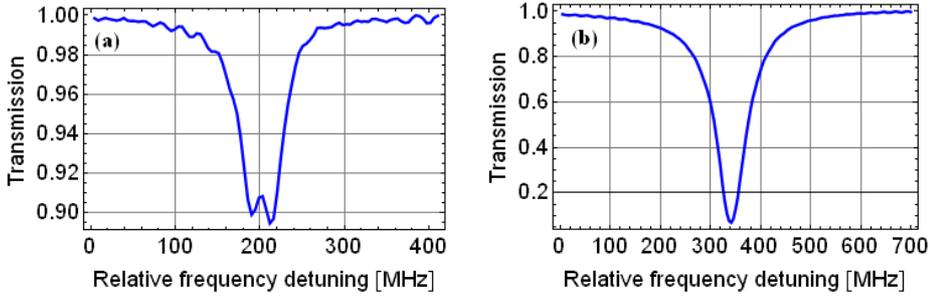

Fig. 2: (a) Normalized transmission spectra of a WGM resonator with loaded $Q = 1.0 \times 10^7$ showing a 26 MHz wide doublet obtained with detector $D_3$. (b) By reducing the loaded quality factor of the resonator to $Q = 3.7 \times 10^6$ the linewidth broadening makes the doublet invisible.

The proof-of-principle experiments reported here have been achieved in air with the set-up shown in Fig. 1 using a widely tunable diode laser at 780nm (New Focus TLB 6312). The microtoroid was fabricated in-house using a 2 µm thick $SiO_2$ layer grown on Si [9] and had major and minor diameters of 60 and 6 µm, respectively. To induce a relatively large mode splitting, we immersed the microtoroid in approximately 100µl of a $10^{-8}$ molar solution of CdSe quantum dots dissolved in N-Decane [10]. The quantum dots had no noticeable absorption at 780nm and might also be replaced by any other scattering microparticle with low absorption.

Dots are attached to the microtoroid as the solvent evaporates. The transmission spectrum of the microtoroid measured with detector $D_3$ is shown in Fig 2a. It has a loaded $Q$ factor of approximately $1.0 \times 10^7$ and features a doublet structure with a splitting of $26 \pm 2$ MHz. As the taper is brought closer to the microtoroid the coupling loss rate $\kappa_1$ is increased, reducing the $Q$ to $3.7 \times 10^6$, and the doublet structure disappears (Fig. 2b). In this regime mode splitting detection is not possible with previously demonstrated techniques. After implementing the interferometer the splitting can still be detected by comparing the spectra measured with detectors $D_1$ and $D_2$. A relative phase shift of $\phi = 0$ or $\pi$ was achieved by using a piezo-electric actuator to move the microtoroid along the taper and to maximize the frequency shift between the two spectra (see eq. 4). One observes that the extremes of the two spectra shown in Fig. 3a are shifted relative to each other by $25 \pm 2$ MHz consistent with the splitting determined at weak coupling (Fig. 2a).

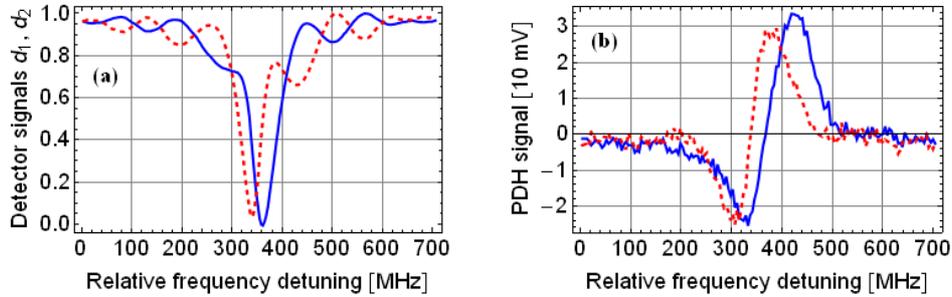

Fig. 3: (a) The minima of the measured detector signals $d_1$ (dotted) and $d_2$ are shifted by $26 \pm 2$ MHz. (b) The zero crossings of the corresponding PDH error signal $e_1$ (dotted) and $e_2$ are also shifted by the same amount and thus constitute an efficient way to measure the splitting with high bandwidth.

As a practical way to detect the mode splitting in real time without continuously scanning the laser and without requiring a fit of the transmission spectrum [3], we derived a Pound Drever Hall (PDH) error signal from each detector [11]. In general the zero crossing of a PDH error signal corresponds to an extremum in the detector signal. A 98 MHz radio frequency signal was applied to an electro-optic modulator, phase modulating the laser. The signals from both detectors were then down modulated at the radio frequency to produce the error signals $e_1$ and $e_2$. Again, the error signals are shifted relative to each other and the frequency difference between their zero crossings corresponds to the splitting (Fig. 3b). Provided that the laser frequency is on or close to resonance, a signal proportional to the splitting can be obtained by subtracting the two PDH error signals. This signal $e_1 - e_2$ has a very high bandwidth, limited only by the modulation frequency of the PDH set-up, i.e. 98 MHz in our demonstration. The laser was locked onto resonance with a PI controller using the combined error signal $e_1 + e_2$.

To conclude, we have demonstrated an interferometric method to detect splitting in microresonators with high sensitivity and high bandwidth even in the case where the splitting is smaller than the linewidth. The proposed technique offers the possibility to investigate single-molecular processes in liquid environments in real-time.

We thank Dr. Mark Fernee, University of Queensland, for providing the quantum dots. The authors gratefully acknowledge the support from the Australian Research Council under Grant No. DP0987146.